\documentclass[12pt,preprint]{aastex}

\begin{document}


\title{On Estimating the Flux of the Brightest Cosmic Ray Source above $57 \times 10^{18}$ eV}


\author{P. W. Younk}
\affil{Colorado State University, Fort Collins, CO 80523}
\email{Patrick.Younk@colostate.edu}



\begin{abstract}
The sources of ultra-high energy cosmic rays are not yet known. However, the discovery of anisotropic cosmic rays above $57 \times 10^{18}$ eV by the Pierre Auger Observatory suggests that a direct source detection may soon be possible. The near-future prospects for such a measurement are heavily dependent on the flux of the brightest source. In this work, we show that the flux of the brightest source above $57 \times 10^{18}$ eV is expected to comprise 10\% or more of the total flux if two general conditions are true. The conditions are: 1.) the source objects are associated with galaxies other than the Milky Way and its closest neighbors, and 2.) the cosmic ray particles are protons or heavy nuclei such as iron and the Greisen-Zatsepin-Kuz'min effect is occurring. The Pierre Auger Observatory collects approximately 23 events above $57 \times 10^{18}$ eV per year. Therefore, it is plausible that, over the course of several years, tens of cosmic rays from a single source will be detected.
\end{abstract}





\section{Introduction}

The field of ultra-high energy cosmic rays has long held more questions than answers. It has been over 40 years since the detection of the first cosmic ray with energy greater than $10^{20}$ eV \citep{Linsley}, and we still do not know where these particles come from. However, we may be on the verge of rapid progress. 

The detection of anisotropic cosmic rays (CR) above $57 \times 10^{18}$ eV \citep{Pierre1, Pierre2} by the newly constructed  Pierre Auger Observatory \citep{Pierre3} has lead to speculation that CR astronomy may soon be a reality. If this is true, it is expected that the brightest source will be the first to be resolved. Since any meaningful astronomical measurement (e.g., sky position, energy spectrum, or morphology) will require the detection of many particles from the source object, the near-future prospects for CR astronomy depend largely on the flux of the brightest CR source in the sky. Consider that the  Auger Observatory detects approximately 23 CR above $57 \times 10^{18}$ eV per year \citep{Pierre2}. If the brightest source contributes 10\% of the flux, the Pierre Auger Observatory will detect ~23 events from this source in a decade; the era of CR astronomy will begin. If, on the other hand, the flux of the brightest source is far below 1\%, the Pierre Auger Observatory may not detect more than one CR from this source during its 20-year operational lifetime.

In this letter, we estimate the flux of the brightest source above $57 \times 10^{18}$ eV by postulating two general conditions. The conditions are consistent with a large number of leading source scenarios; e.g., radio loud galaxies \citep{Fraschetti}, active galactic nuclei \citep{Biermann, Farrar}, and gamma ray bursts \citep{Waxman, Wick}.
The first condition is that the source objects are associated with galaxies, but the sources are relatively rare or rarely active such that many galaxies, such as our own and our closest neighbors, do not contain a luminous source. This implies a source number density $dN/dV < 10^{-2} \quad\mbox{Mpc}^{-3}$. The second condition is that the CR particles are protons or heavy nuclei such as iron and the Greisen-Zatsepin-Kuzmin (GZK) energy loss processes \citep{Greisen,Zatsepin} are occurring. Above $10^{20}$ eV, the energy loss length   for protons and iron is tens of Mpc. This effectively limits the propagation distance of the highest energy CR to 100 Mpc, which implies a source number density $dN/dV > 10^{-6} \quad\mbox{Mpc}^{-3}$. The energy loss lengths of intermediate weight nuclei are less than either protons or iron-like heavy nuclei. Therefore, the second condition is conservative, i.e. it implies the broadest range of source number densities. This range of source densities fully encompasses the ranges recently estimated by \citet{Takami} and \citet{Cuoco}.

\section{Simplified Analytical Treatment}


In this work, we only consider the integrated CR flux above a threshold energy $E_{th} = 57 \times 10^{18}$ eV. Because of GZK energy losses, sources beyond a distance $D$, the so-called GZK horizon, will be nearly undetectable in this energy range. The GZK horizon is a strong function of the threshold energy, decreasing as $E_{th}$ increases. By an accident of nature, the GZK horizon is approximately the same for both protons and iron nuclei \citep{Harari}. Therefore, we will only examine the proton case in detail here, realizing that the iron case will have similar results.

\begin{figure}
\epsscale{1.0}
\plotone{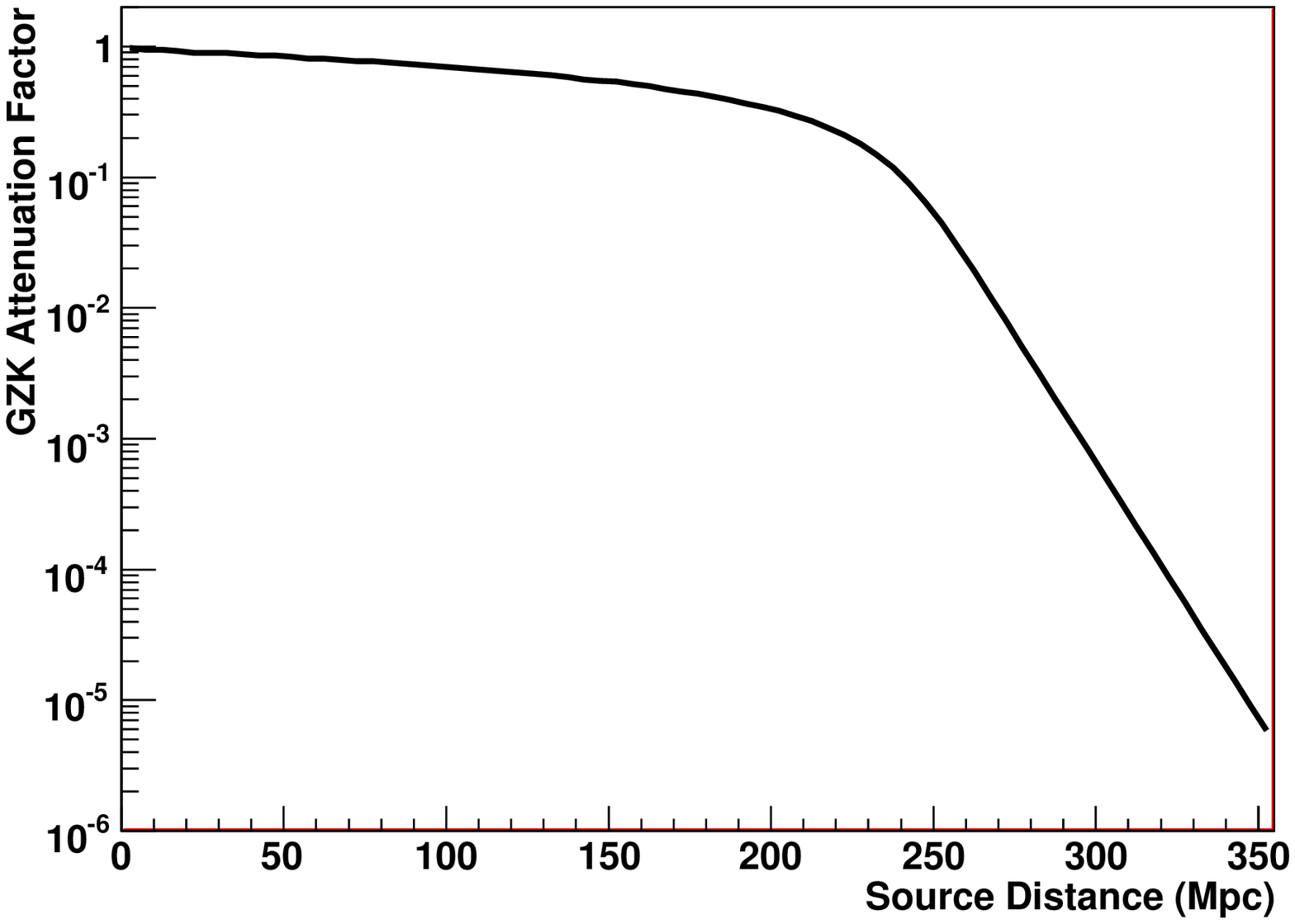}
\caption{The GZK attenuation factor of a proton source observed in an energy range $E > 57 \times 10^{18}$ eV. We assume that the near-source energy spectrum is a power law $dn/dE \propto E^{-2.6}$.}
\label{fig:Attenuation}
\end{figure}

To reproduce the observed energy spectrum we assume that the proton injection spectrum (i.e., the near-source spectrum) is a power law $dn/dE \propto E^{-\alpha}$ with $\alpha = 2.6$, as suggested by \citet{Allard}. Our estimation of the GZK horizon then proceeds as follows. As in \citet{Harari}, we define a GZK attenuation factor $A(E_{th}, x)$ as the fraction of particles originally above $E_{th}$ that still have an energy above $E_{th}$ after traversing a distance $x$. The attenuation factor is then $A(E_{th},x) = ((E_{th} + \Delta E)/E_{th})^{1-\alpha}$, where $\Delta E$ is the average energy loss of a proton traversing a distance $x$. The energy loss $\Delta E$ is calculated by solving $dE/dx = -E/ \lambda$, where $\lambda$ is the proton energy loss length in the extragalactic medium (e.g., see \citet{Protheroe}). This is the so-called continuous energy loss approximation. In Fig.~\ref{fig:Attenuation}, we show the results of this calculation for $E_{th} = 57 \times 10^{18}$ eV. There is a rather sharp breakpoint at approximately 250 Mpc beyond which the attenuation factor rapidly decreases exponentially. A reasonable choice for the GZK horizon is then 250 Mpc.

Let us examine a simplified scenario where there are $N$ sources of equal luminosity $L$ evenly distributed within a sphere of radius $D = 250$ Mpc centered at earth. We approximate the attenuation factor as a step function, with a value of one within the sphere and zero outside. The flux received at the detector from a single source inside the sphere is $q = \omega L r^{-2}$, where $\omega$ is the detector's sky exposure at the source location, and $r$ is the source distance. We assume $\omega$ is constant over all sky locations; i.e., the detector sees equally well over the entire celestial sphere. 

Because the sources are distributed evenly, the number of sources in a shell of thickness $dr$ is
\begin{equation}
dN = 3Nr^{2}dr/D^{3} ,
\label{eq:RadialDensity}
\end{equation}
and the total flux is $q_{tot} = \int_{0}^{D}  qdN = 3 \omega LN/D^{2}$. We find the median distance to the nearest source $\widetilde{R}$ by solving $\int_{0}^{\widetilde{R}} dN = 1/2$, which gives $\widetilde{R} = D(2N)^{-1/3}$. Then the expected flux of the nearest source relative to the total flux, is
\begin{equation}
\bar{Q} \approx \omega L \widetilde{R}^{-2} / q_{tot} = (27N/4)^{-1/3} .
\label{eq:AnalyticQ}
\end{equation}
Since $N \propto \rho D^{3}$, where $\rho$ is the density of sources, Eq.~\ref{eq:AnalyticQ} implies $\bar{Q} \propto \rho^{-1/3} D^{-1}$. Because of the quadratic form of Eq.~\ref{eq:RadialDensity}, the median distance   is an overestimate of the mean (expected) distance. Therefore, Eq.~\ref{eq:AnalyticQ} is a lower limit of the expected flux of the brightest source. For a source density of $10^{-6} \quad\mbox{Mpc}^{-3}$, Eq.~\ref{eq:AnalyticQ} gives $\bar{Q} = 13\%$; while for a source density of $10^{-2} \quad\mbox{Mpc}^{-3}$, $\bar{Q} = 0.6\%$.




\section{Numerical Treatment}

In the following numerical treatment, we calculate $\bar{Q}$ directly via a Monte Carlo algorithm with several refinements over the analytical treatment: 1.) the GZK attenuation function $A$ is a smooth function of source distance (not a step function), 2.) the sky exposure $\omega$ is not constant over the entire sky, 3.) the source luminosities are not necessarily all equal, and 4.) the distribution of sources is not homogeneous. These refinements tend to increase $\bar{Q}$ above the estimate given in Eq.~\ref{eq:AnalyticQ}.

We calculate the flux from each source as $q = \omega A L r^{-2}$, the attenuation factor $A$ being a smooth function of source distance as calculated in the previous section, and the sky exposure $\omega$ being a function of the source declination. We use the $\omega$ function developed by \citet{Sommers} with a detector at latitude $-35^{\circ}$ (the location of the Southern site of the Auger  Observatory), and a maximum zenith angle acceptance of $60^{\circ}$ (a standard quality cut implemented by the Pierre Auger Collaboration).

We describe the source luminosity distribution with a differential luminosity function $\phi = dN/dL$. We consider two scenarios. The first scenario is that $\phi$ is a power law $\phi \propto L^{-2}$ with $L$ ranging over three orders of magnitude. This scenario is motivated by the fact that the luminosity function of many astronomical objects (stars, normal galaxies, and AGN) is a power law over a similar range of luminosities; e.g., see \citet{Basu, Xu, Mauch}. The simple ubiquity of the power law form makes this scenario worthy of consideration. The second scenario is the same as in the analytical treatment, that all sources have equal luminosity.

One of our starting assumptions was that the sources are associated with galaxies. The distribution of galaxies on Mpc scales is far from homogeneous. Therefore, the number of sources in a shell of radius $r$ is not necessarily well approximated by Eq.~\ref{eq:RadialDensity}. The largest relative deviations from Eq.~\ref{eq:RadialDensity} occur at short distances. For example, the number of bright galaxies ($M_B < -19.5$) within 100 Mpc is approximately 23,000 \citep{Marzke}. From this, Eq.~\ref{eq:RadialDensity} predicts that there are 0.023 bright galaxies within 1 Mpc and 2.9 within 5 Mpc. However, the observed numbers are 2 and 8, respectively \citep{Tully}. To consider this local over-density in a simple way, we use a modified radial density function:
\begin{equation}
dN/dr =
\cases{ 
30 N r/D^3 & : $r \leq 10~\mbox{Mpc}$ \cr
3Nr^2/D^3 & : $r > 10~\mbox{Mpc.}$}
\label{eq:RadialDensity2}
\end{equation}
We do not try to replicate the directional distribution of galaxies since this has no bearing on the results.

Our numerical treatment calculates $\bar{Q}$ by averaging the results of many Monte Carlo realizations. We form a single Monte Carlo realization as follows. $N$ sources are assigned a random sky position, distance, and luminosity. The sky positions are chosen to follow a flat distribution (each sky position has equal probability of containing a source), the distances are chosen to follow the radial density function given by Eq.~\ref{eq:RadialDensity2} out to a maximum distance of 350 Mpc, and the luminosities are chosen to follow either a power law distribution (scenario 1) or to be all equal (scenario 2). The flux $q$ from each source is computed, then the relative flux of the brightest source is computed as $Q = \max\{ q_1,\ldots,q_N \} / \mbox{sum} \{ q_1,\ldots,q_N \}$. We calculate $\bar{Q}$ by averaging over 1000 realizations. This is repeated for several different source densities.

We show the results of this analysis in Fig.~\ref{fig:Qbar}. The error bars represent the 10-90\% quantiles of each set of realizations. At a source density of $10^{-6} \quad\mbox{Mpc}^{-3}$, the expected value of $\bar{Q}$ for scenario 1 is 40\% and the 10-90\% quantile range is 19\% to 70\%. At a source density of $10^{-2} \quad\mbox{Mpc}^{-3}$, the expected value of $\bar{Q}$ is 10\% and the 10-90\% quantile range is 1.8\% to 26\%.

The numeric results are greater than the analytical treatment at all source densities. This is expected since the analytic treatment is a lower limit. From a source density of $10^{-6} \quad\mbox{Mpc}^{-3}$ to $10^{-4} \quad\mbox{Mpc}^{-3}$, the numerical results follow the general trend of the analytic treatment: decreasing by a factor of approximately 4 as the source density increases by a factor of 100.  At source densities greater than $10^{-4} \quad\mbox{Mpc}^{-3}$, the numerical results flatten (i.e., they are nearly independent of source density). This is the source density above which we expect the closest sources to be within 10 Mpc and the linear region of Eq.~\ref{eq:RadialDensity2} to be important. If no local overdensity is assumed, i.e. Eq.~\ref{eq:RadialDensity} is substituted for Eq.~\ref{eq:RadialDensity2}, then the numerical results do not flatten. In this case, the expected value of $\bar{Q}$ for scenario 1 at a source density of $10^{-2} \quad\mbox{Mpc}^{-3}$ is 4.0\%.

The relatively weak dependence of $\bar{Q}$ on source density is caused by two effects which somewhat balance each other. The first is that a greater number of sources will tend to diminish the relative flux of the brightest source by increasing the background. The second is that a greater number of sources will tend to increase the flux of the brightest source since there is a greater probability of a source being relatively close or luminous or both. For the special case $dN/dr \propto r$, the tendencies exactly balance such that $\bar{Q}$ is independent of source density.

The weak dependence of $\bar{Q}$ on source density implies that estimates of source density based solely on event clustering will have a large uncertainty. Indeed, the recent CR source density estimates by \citet{Takami} and \citet{Cuoco} reflect this.

In general, scenario 1 has a greater $\bar{Q}$ than scenario 2, however the difference is well within the 10-90\% quantile range. That is, the results are not sensitive to the details of the source luminosity function. Regardless of this, scenario 1 (power-law luminosity function) is the more appropriate and the more general consideration. For instance, in scenario 1 the brightest sources are spread over a relatively large range of distances, whereas in scenario 2 the brightest sources are always clustered relatively nearby. If scenario 2 is adopted, a lack of events from a nearby collection of matter can be used to argue against a model with relatively high source densities (e.g., see \citet{Gorbunov}). Such a constraint is the result of attributing a difference of magnitude to a difference in type, and is almost certainly artificial.

\section{Prospects for CR Astronomy}

\begin{figure}
\epsscale{1.0}
\plotone{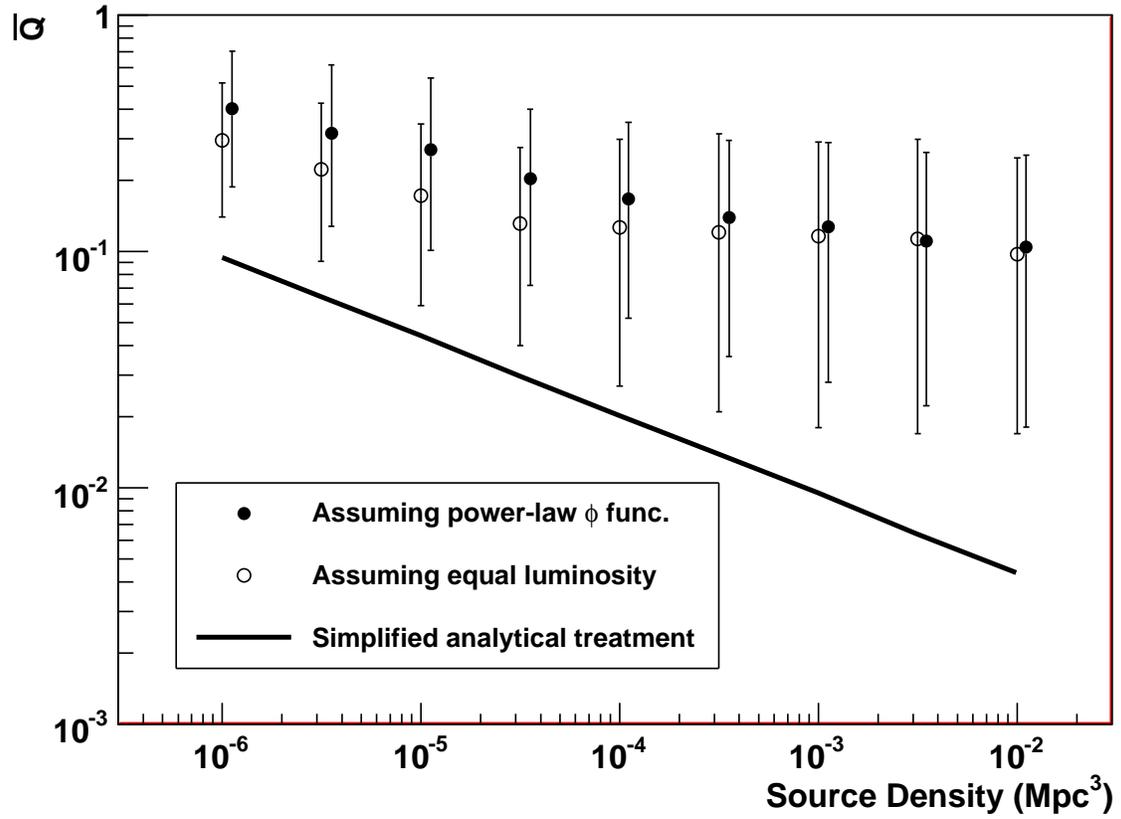}
\caption{The expected flux, relative to the total flux, of the brightest cosmic ray source above $57 \times 10^{18}$ eV as a function of source number density. The points are slightly offset on the x-axis for clarity. The analytic approximation is Eq.~\ref{eq:AnalyticQ}.}
\label{fig:Qbar}
\end{figure}

We expect that the flux at earth from any particular extragalactic CR source is nearly constant over a time span of several years. Even if the particles are accelerated in a rather short burst, the CR pulse is broadened in time because of magnetic deflections, with the highest energy particles arriving first. \citet{Waxman} has shown that a time broadening of 100 years or more for a 100 Mpc distant source is a reasonable expectation based on our present knowledge of extragalactic magnetic fields.
If source persistence is indeed due to extragalactic magnetic fields, the apparent source density and apparent source luminosity are functions of these intervening magnetic fields, i.e. they are functions of distance. In this case, the local apparent source density should be used with Fig.~\ref{fig:Qbar} since we are only concerned with the flux of the brightest source which is typically one of the nearest.

If we assume the flux from the brightest source is steady and $\bar{Q}$ is 10\%, then we expect the Pierre Auger Observatory to acquire 2.3 events from this source per year. The angular size of the source is dependent on the magnetic rigidity of the particle and the intervening magnetic fields, both of which are not well constrained. However, even if the source covers a $20^{\circ} \times 20^{\circ}$ area (approximately 1\% of the sky observed by the Pierre Auger Observatory), the number of events from this region will be ten times the expected amount from an isotropic background. In this respect, a source detection is expected to occur in a matter of years. 

In this work, we have concentrated on the energy range above $57 \times 10^{18}$ eV. However, if the extragalactic source paradigm is correct, CR astronomy is possible down to the energy where the diffuse galactic CR become dominant over the extragalactic CR (most likely somewhere below $10^{19}$ eV \citep{Allard}). At lower energies, $\bar{Q}$ will be less (because as the energy threshold $E_{th}$ decreases the GZK horizon $D$ increases, and $\bar{Q}$ is inversely proportional to $D$) and the angular size of the sources will be larger because of decreasing magnetic rigidity.

\acknowledgments
The author thanks his fellow Pierre Auger collaborators, and acknowledges the support of the National Science Foundation (award number 0838088) and the Michigan Space Grant Consortium.

\clearpage



\clearpage






\end{document}